\documentclass[preprint,onecolumn,nofootinbib]{revtex4}
\usepackage[colorlinks=true,linkcolor=blue,urlcolor=blue,filecolor=black,citecolor=red,pdfstartview=FitV,pdftitle={},pdfsubject={},pdfkeywords={},pdfpagemode=None,bookmarksopen=true]{hyperref}
\usepackage{graphicx}
\usepackage{amsmath}
\usepackage{amsfonts}
\usepackage{amssymb,ulem}
\usepackage{color,xcolor}%
\usepackage{CJK}
\usepackage{subfigure}
\usepackage{amsthm,amsmath,amssymb}
\usepackage{mathrsfs}
\usepackage{textcomp}

\usepackage{dcolumn}
\usepackage{float}
\begin{document}
\title{Charge transport properties in a novel holographic quantum phase transition model}

\author{Guoyang Fu$^{1}$}
\thanks{FuguoyangEDU@163.com}
\author{Huajie Gong$^{1}$}
\thanks{huajiegong@qq.com}
\author{Peng Liu $^{2}$}
\thanks{phylp@email.jnu.edu.cn}
\author{Xiao-Mei Kuang$^{1}$}
\thanks{xmeikuang@yzu.edu.cn}
\author{Jian-Pin Wu$^{1}$}
\thanks{jianpinwu@yzu.edu.cn, corresponding author}
\affiliation{
	$^1$ Center for Gravitation and Cosmology, College of Physical Science and Technology, Yangzhou University, Yangzhou 225009, China }
\affiliation{$^2$ Department of Physics and Siyuan Laboratory, Jinan University, Guangzhou 510632, P.R. China}

\begin{abstract}

We investigate the features of charge transport in a novel holographic quantum phase transition (QPT) model with two metallic phases: normal metallic and novel metallic. The scaling behaviors of direct current (DC) resistivity and thermal conductivity at low temperatures in both metallic phases are numerically computed. The numerical results and the analytical ones governed by the near horizon geometry agree perfectly.
Then, the features of low-frequency alternating current (AC) electric conductivity are systematically investigated.
A remarkable characteristic is that the normal metallic phase is a coherent system, whereas the novel metallic phase is an incoherent system with non-vanishing intrinsic conductivity. Especially, in the novel metallic phase, the incoherent behavior becomes stronger when the strength of the momentum dissipation enhances.

\end{abstract}

\maketitle

\section{Introduction}

AdS/CFT correspondence relates a weakly coupled gravitational theory to a strongly coupled quantum field theory without gravity in the large N limit \cite{Maldacena:1997re,Gubser:1998bc,Witten:1998qj,Aharony:1999ti}. This duality provides some physical insight into the associated mechanisms of the strongly coupled quantum many-body systems. With the help of AdS/CFT correspondence, significant progresses have been made in understanding novel mechanisms for superconductivity \cite{Hartnoll:2008vx} and metal-insulator phase transition (MIT) \cite{Donos:2012js,Ling:2014saa,An:2020tkn}, transport properties \cite{Hartnoll:2009sz,Natsuume:2014sfa,Hartnoll:2016apf,Baggioli:2019rrs,Baggioli:2021xuv}, entanglement entropy \cite{Ryu:2006bv,Takayanagi:2012kg,Lewkowycz:2013nqa,Hubeny:2007xt,Dong:2016hjy}, quantum chaos \cite{Kitaev:2014,Maldacena:2015waa,Roberts:2016wdl}, and so on.

In holography, the phase is essentially depicted by geometry, such that phase transition is characterized by the transition of geometry  \cite{Donos:2012js,Donos:2014uba,Fu:2022qtz}. Specifically, the lattice operator in holographic model induces infrared (IR) instability, which leads to a new IR fixed point. The shift between different IR fixed points results in a phase transition. There is also another mechanism driving phase transition. It is the strength of lattice deformation that gives rise to some kind of bifurcating solution such that phase transition happens \cite{Donos:2012js}.

MIT, as a prominent example of quantum phase transition (QPT), have been implemented and widely explored from holography \cite{Donos:2012js,Donos:2014oha,Donos:2013eha,Donos:2014uba,Ling:2014saa,Baggioli:2014roa,Kiritsis:2015oxa,Ling:2015epa,Ling:2015exa,Ling:2016dck,Mefford:2014gia,Baggioli:2016oju,Andrade:2017ghg,Bi:2021maw,An:2020tkn,Fu:2022qtz}. It is identified by the transition on the sign of slope of the DC (direct-current) conductivity $\sigma_{DC}$ near extremely low temperature $T$. Specifically, $\partial_T\sigma_{DC}<0$ indicates a metallic phase, while $\partial_T\sigma_{DC}>0$ demonstrates an insulating phase, and $\partial_T\sigma_{DC}=0$ describes the critical point transiting between the metallic phase and insulating phase.

Recently, we find interesting phenomena in the holographic EMDA (Einstein-Maxwell-dilaton-axions) model \cite{Donos:2014uba,Fu:2022qtz} that for certain model parameter $\gamma$, when we change the lattice parameters, the system exhibits consistent temperature behavior of DC conductivity but with two different IR geometries. According to the geometry viewpoint \cite{Donos:2012js}, we argue that there is a novel holographic QPT \cite{Fu:2022qtz}. We refer to the phase with AdS$_2 \times \mathbb{R}^2$ IR geometry as the normal metallic phase and the phase with non-AdS$_2 \times \mathbb{R}^2$ (hyperscaling violation) geometry as the novel metallic phase \cite{Fu:2022qtz}. It would be interesting to further study the holographic properties of this novel holographic QPT model. Thus, this paper proposes to investigate the charge transport properties by studying DC resistivity and the AC (alternating current) conductivity of the novel state.

We are specially interested in the behavior of the AC conductivity at low frequency as it can indicate the metal in a coherent phase or an incoherent one. Usually, for a coherent metallic phase, the behavior of the AC conductivity at low frequency is fitted by the standard Drude formula
\begin{eqnarray}\label{Drude formula}
	\sigma(\omega)=\frac{\sigma_{DC}}{1-i \omega\tau},
\end{eqnarray}
with $\tau$ the relaxation time, which depends on the temperature, and if the above formula is violated, the corresponding metal could be in an incoherent phase. In holographic framework, it is common that the dual metal phase could transit between coherent and incoherent phases. For instance, in the Einstein-Maxwell-axions (EMA) theory or Gubser-Rocha-axions model \cite{Davison:2015bea,Zhou:2015qui}, the standard Drude formula \eqref{Drude formula} was found to be satisfied only when momentum dissipation of system is weak. But as the momentum dissipation enhances, the low-frequency behavior of AC conductivity cannot be fitted by \eqref{Drude formula}, implying that the metal is in incoherent phase, which is depicted by the modified Drude formula \cite{Davison:2015bea,Zhou:2015qui}
\begin{eqnarray}\label{Holographic modify formula}
	\sigma(\omega)=\frac{\sigma_{DC}-\sigma_Q}{1-i \omega\tau}+\sigma_Q,
\end{eqnarray}
where $\sigma_Q$ is known as the intrinsic conductivity \cite{Davison:2015taa}. From the modified Drude formula \eqref{Holographic modify formula}, the low-frequency conductivity for incoherent metallic phases attributes to two compositions: the coherent contribution due to momentum relaxation and the incoherent contribution due to the intrinsic current relaxation. Since we could expect that the charge transport behaviors in the normal and novel metallic phases we found in holographic EMDA model \cite{Fu:2022qtz} could be different, the careful study on the AC conductivity at low frequency could help to further understand the normal and novel phases. 

We organize the paper as follows. In section \ref{sec-EMDA-model}, we review the holographic EMDA model proposed in \cite{Donos:2014uba,Fu:2022qtz}. Then, in section \ref{sec-dcres}, we calculate the DC resistivity and DC thermal conductivity and study its scaling behavior at low temperatures. Section \ref{sec-ac} is dedicated to the properties of low-frequency AC conductivities. Finally, in section \ref{sec-con}, we conclude the paper by presenting a summary and a discussion.

\section{Holographic setup}\label{sec-EMDA-model}

In this section, we briefly review the holographic setup of EMDA model. For the details, please refer to Refs.\cite{Donos:2014uba,Fu:2022qtz}.
The EMDA action we consider is \cite{Donos:2014uba,Fu:2022qtz}
\begin{eqnarray}\label{EMDA-Action}
S= {}\int d^{4}x \sqrt{-g} \left[ R +6 \cosh\psi - \frac{3}{2} [ (\partial \psi)^2+4\sinh^2\psi (\partial{\chi})^2 ] - \frac{1}{4} \cosh^{\gamma /3}(3\psi)F^2 \right]\,,
\end{eqnarray}
where we have fixed the $AdS$ radius $L=1$. $\psi$ is the dilaton field coupled with the Maxwell field $F\equiv dA$ and the axion field $\chi$ and $\gamma$ is the coupling parameter. The system exhibits attractive QPT landscapes depending on this coupling parameter, as described in Refs. \cite{Donos:2014uba,Fu:2022qtz}. We are interested in the novel QPT, which changes from a normal metallic phase to a novel metallic one when $\gamma$ in the region of $\gamma>3$. Without loss of generality, we set $\gamma=9/2$ throughout this paper.

To solve this holographic system \eqref{EMDA-Action}, we assume the following ansatz
\begin{eqnarray} \label{metric}
ds^2&&=\frac{1}{z^2}\big{[}-(1-z)p(z)U(z)dt^2+\frac{dz^2}{(1-z)p(z)U(z)}+V_1(z) dx^2+V_2(z) dy^2\big{]}, \\ \nonumber
A&&=\mu(1-z)a(z) dt,  \\ \nonumber
\psi&&=z^{3-\Delta}\phi(z), \\ \nonumber
\chi&&=\hat{k} x,
\end{eqnarray}
where $p(z)=1+z+z^2-\mu^2 z^3/4$, $\hat{k}$ is the charge of the axion field, which depicts the strength of momentum dissipation, and $\mu$ is interpreted as the chemical potential. In our model \eqref{EMDA-Action}, the conformal dimension of the dilaton field is $\Delta=2$.

The action \eqref{EMDA-Action} with the ansatz \eqref{metric} produces
four second order  ordinary differential equations (ODEs) for $V_1\,, V_2\,, a\,, \phi$ and one first order ODE for $U$. The asymptotic AdS$_4$ on the conformal boundary requires that
\begin{eqnarray}
	U(0)=1\,, \ V_1(0)=1\,, \ V_2(0)=1\,, \ a(0)=1\,, \ \phi(0)=\hat{\lambda}\,,
\end{eqnarray}
where $\hat{\lambda}$ is the source of the dilaton field operator in the dual field theory and characterizes the lattice deformation. Collecting all the above information with the regular boundary conditions at the horizon, we can solve this holographic system numerically.

The Hawking temperature of black hole is
\begin{eqnarray}
	\hat{T}=\frac{(12-\mu^2)U(1)}{16 \pi}\,,
\end{eqnarray} 
which is considered as the temperature of the dual theory.
We set the chemical potential $\mu$ as the scaling unit as our previous work \cite{Fu:2022qtz} and this system is described by
the three dimensionless parameters $\{T , \lambda , k \} \equiv \{\hat{T}/\mu , \hat{\lambda}/\mu , \hat{k}/\mu \}$.

\section{DC electric and thermal conductivities}\label{sec-dcres}

In this section, we shall mainly investigate the properties of electric and thermal conductivities over the novel QPT. Following
the scheme proposed in \cite{Donos:2014uba} (also see \cite{Blake:2014yla,Donos:2014cya,Ling:2016dck}) we can derive the DC resistivity\footnote{Here, it would be clearer and more convenient to discuse the DC resistivity, which is the reciprocal of the DC electric conductivity.} and the thermal conductivity, which are given by
\begin{eqnarray}\label{dc resistivity}
	&&
	\rho=\left(\sqrt{\frac{V_2}{V_1}}\left(\cosh^{\gamma/3}(3\phi)+\frac{V_1 a^2 \cosh^{2/3\gamma}(3\phi)}{12 k^2 \sinh^2(\phi)}\right)\right)^{-1}\bigg{|}_{z \to 1}\,, \\
	&&
	\bar{\kappa}=\frac{4\pi^2 \sqrt{V_1 V_2 }T}{3 k^2 \sinh^2(\phi) }\bigg{|}_{z \to 1}\,.
\end{eqnarray}
For the detailed derivation, please refer to \cite{Donos:2014uba,Blake:2014yla,Donos:2014cya,Ling:2016dck}. After numerically obtaining the background solution, we can then determine the DC resistivity and DC thermal conductivity using the aforementioned expressions.

\begin{figure}[H]
	\centering
	\includegraphics[width=0.48\textwidth]{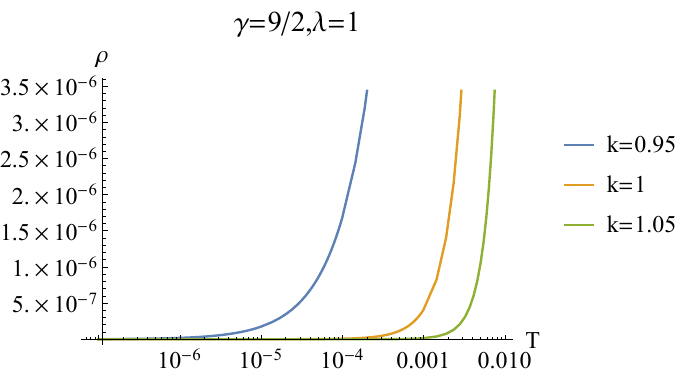}\hspace{0.5mm}	
	\includegraphics[width=0.48\textwidth]{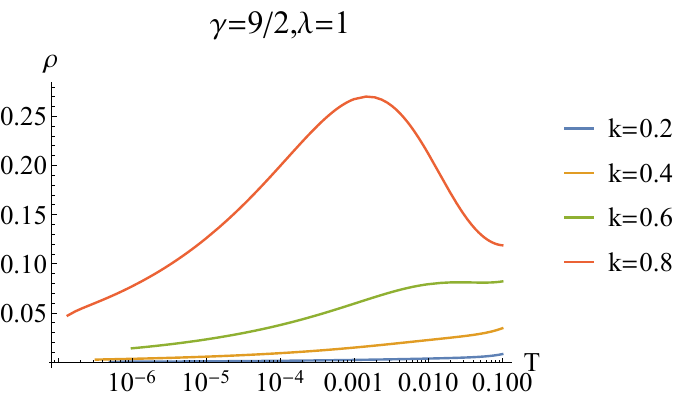}
	\caption{Semilogarithmic plots of DC resistivity $\rho$ as a function of the temperature for different $k$. Here we fix $\gamma=9/2$ and $\lambda=1$. The left plot is for the normal metallic phase, while the right one is the novel metallic phase.}
	\label{dc resistivity with k}
\end{figure}
Fig.\ref{dc resistivity with k} shows the DC resistivity $\rho$ as a function of the temperature for different $k$. The left plot is the normal metallic phase, while the right one is the novel metallic phase. As expected, for the normal metallic phase, the DC resistivity decreases with the temperature. 

In the novel metallic phase, the results are completely different. When $k$ is small, the DC resistivity decreases with the temperature, which is obviously metallic behavior. While for large $k$ (for example, $k=0.8$, see the right plot in Fig.\ref{dc resistivity with k}), as the temperature decreases, the DC resistivity increases at first, indicating an insulating behavior, and then decreases, indicating a metallic behavior. However, at extremely low temperature, this holographic system indeed exhibits metallic behavior. 

\begin{figure}[H]
	\centering
	\includegraphics[width=0.48\textwidth]{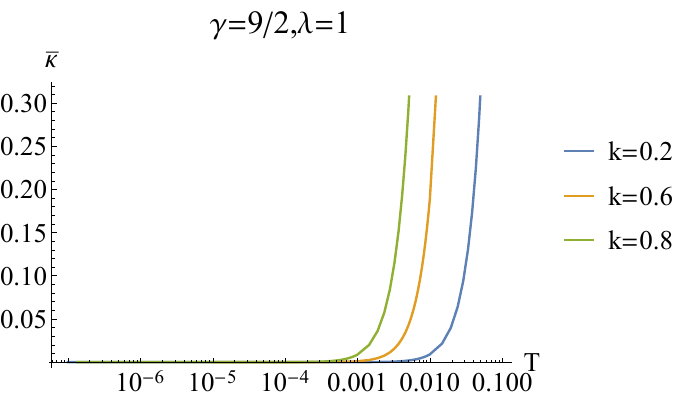}
	\caption{Semilogarithmic plots of DC thermal conductivity $\bar{\kappa}$ as a function of the temperature for different $k$in the novel metallic phase. Here we fix $\gamma=9/2$ and $\lambda=1$. }
	\label{thermal-k_HV}
\end{figure}
\begin{figure}[H]
	\centering
	\includegraphics[width=0.48\textwidth]{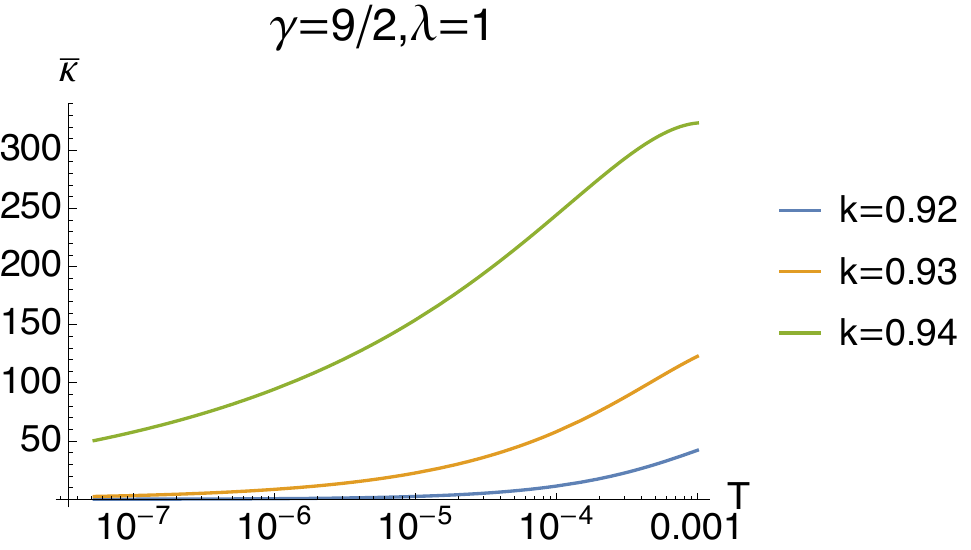}\hspace{0.5mm}	
	\includegraphics[width=0.48\textwidth]{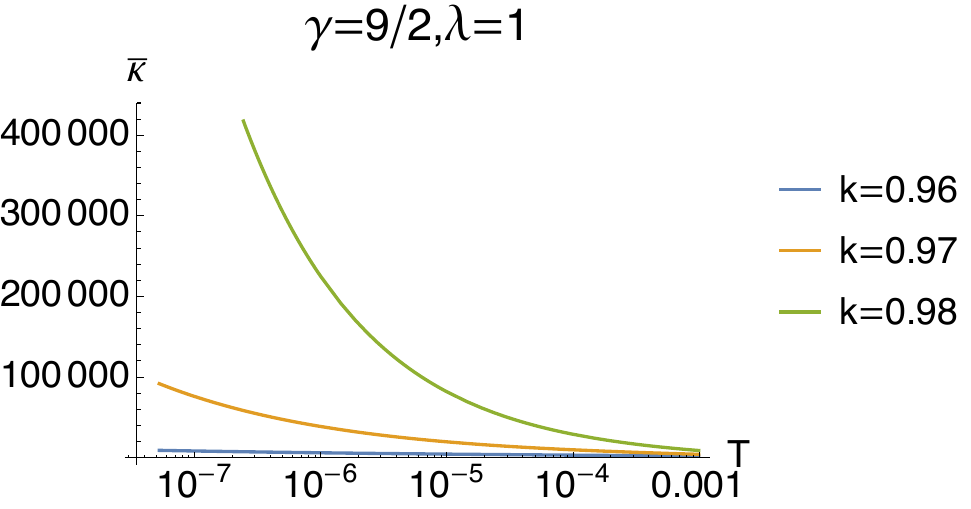}
	\caption{Semilogarithmic plots of DC thermal conductivity $\bar{\kappa}$ as a function of the temperature for different $k$ in the normal metallic phase. Here we fix $\gamma=9/2$ and $\lambda=1$.}
	\label{thermal-k_AdS2}
\end{figure}

We now turn our attention to the DC thermal conductivities. In the novel metallic phase, the thermal conductivity $\bar{\kappa}$ decreases as temperature decreases and eventually approaches zero at low temperatures (Fig.\ref{thermal-k_HV}), suggesting a thermal insulating behavior.

In the normal metallic phase, the DC thermal conductivity shows different behaviors depending on the lattice parameters. As the system approaches the QCP, the DC thermal conductivity $\bar{\kappa}$ behaves similarly to the novel metallic phase, i.e., it decreases with decreasing temperature (shown in the left plot of Fig. \ref{thermal-k_AdS2}). However, when the system is away from the QCP, $\bar{\kappa}$ increases with decreasing temperature, eventually diverges as the temperature approaches zero (as shown in the right plot of Fig. \ref{thermal-k_AdS2}). These findings are further reinforced by the scaling behaviors, which are demonstrated below.
 
To validate our numerical results, we extract the scaling behavior of the DC resistivity and the DC thermal conductivity at low temperature from our numerical results and compare it with the analytical results obtained in Refs.\cite{Donos:2014uba,Donos:2014cya} that are governed by the near horizon geometry.

As discussed in the \cite{Donos:2014uba,Fu:2022qtz}, the model \eqref{EMDA-Action} exhibits two types of IR fixed point solutions as $T\to 0$: the AdS$_2\times \mathbb{R}^2$ fixed point and a hyperscaling violation fixed point. The scaling behaviors of the DC resistivity $\rho$ and the thermal conductivity $\bar{\kappa}$ at low temperatures for the AdS$_2\times \mathbb{R}^2$ solution are given by \cite{Donos:2014cya}
\begin{eqnarray}\label{AdS2_scale}
\rho \sim T^{2\Delta(k)-2}\,, \ \ \bar{\kappa}\sim T^{3-2\Delta(k)}\,,
\end{eqnarray}
where $\Delta(k)$ is the scaling dimension of the dilaton field and it is given by
\begin{eqnarray}\label{Delta_k}
\Delta(k)=\frac{1}{2}+\frac{1}{6} \sqrt{24 e^{-2 v_{10}} k^{2}-3(12 \gamma+1)}\,.
\end{eqnarray}
The scaling dimension $\Delta(k)$ obviously depends on both the charge of the axion field $k$ and the constant $v_{10}$, which is determined by the IR geometry data and must be computed numerically. It is worth noting that the coupling function of the kinetic term of the axion field $\chi$ follows a scaling behavior $\sim T^{2\Delta(k)-2}$ in the limit of zero temperature \cite{Donos:2014cya}. To avoid divergence as $T$ approaches $0$, it is necessary for $\Delta(k)$ to be greater than 1, i.e., $\Delta(k)>1$.

According to Eq. \eqref{AdS2_scale}, the DC resistivity $\rho$ tends to zero as a temperature $T$ approaches zero, regardless of whether the system is in a normal metallic or novel metallic phase, as indicated by the condition $\Delta(k)>1$. This result is in agreement with the left plot of Fig.\ref{dc resistivity with k}.
However, the behavior of the thermal conductivity $\bar{\kappa}$ is quite distinct and depends on $\Delta(k)$. Specifically, when $1<\Delta(k)<3/2$, $\bar{\kappa}$ tends to zero; when $\Delta(k)=3/2$, $\bar{\kappa}$ remains constant; and when $\Delta(k)>3/2$, $\bar{\kappa}$ diverges. These observations are consistent with the findings depicted in Figs. \ref{thermal-k_HV} and \ref{thermal-k_AdS2}.

\begin{figure}[H]
	\centering
	\includegraphics[width=0.48\textwidth]{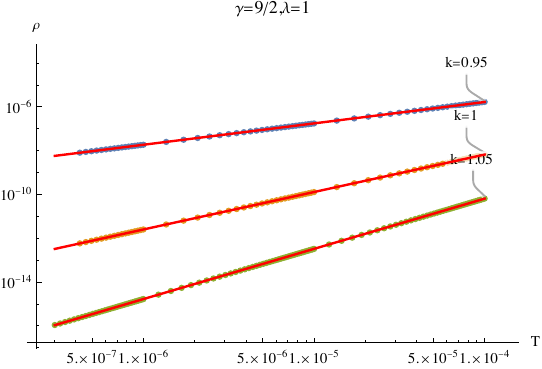}\hspace{0.3mm}
	\includegraphics[width=0.48\textwidth]{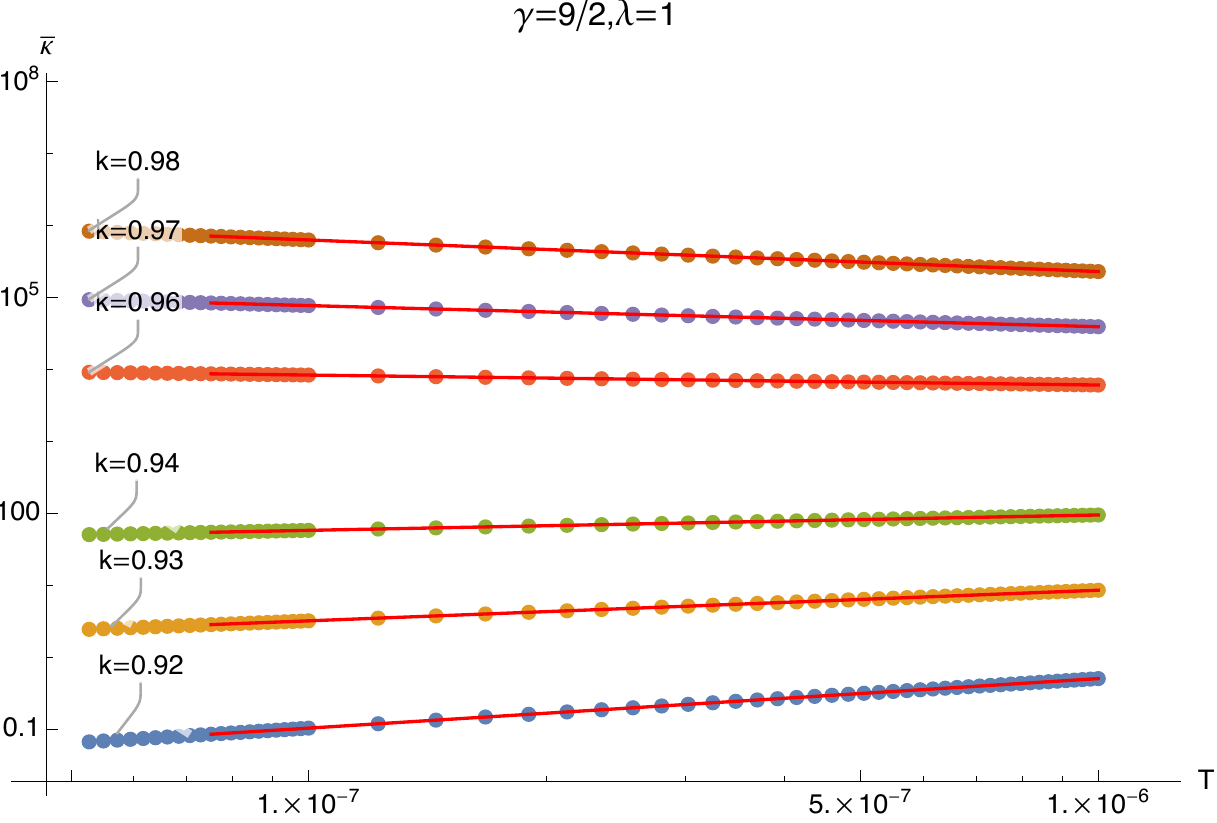}
	\caption{The log-log plot of the DC resistivity $\rho$ and thermal conductivity $\bar{\kappa}$ as the function of temperature $T$ for different $k$ in the normal phase. The numerical results are represented by dots, while the analytical results, evaluated using Eq.\eqref{Delta_k}, are depicted as solid lines.
	}
	\label{AdS2_scaling_relationship}
\end{figure}

In Fig.\ref{AdS2_scaling_relationship}, we present the log-log plot of the DC resistivity $\rho$ and the DC thermal conductivity $\bar{\kappa}$ as a function of temperature $T$ in the normal metallic phase. The numerical results are represented by dots, while the analytical results, evaluated using Eq.\eqref{Delta_k}, are depicted as solid lines. These plots demonstrate that the numerical results are in excellent agreement with the analytical results. To quantify our findings, we performed a numerical fit of the data using the form given in Eq.\eqref{AdS2_scale}. The scaling dimension $\Delta(k)$ was determined from the fit and denoted as $\Delta_N(k)$ in Table \ref{ta-0}. In addition, we directly evaluated $\Delta(k)$ using Eq.\eqref{Delta_k}, which we denote as $\Delta_A(k)$ and present in Table \ref{ta-0}. The numerical results are highly consistent with that evaluated in terms of Eq.\eqref{Delta_k}.

\begin{table}[H]
	\centering
	\caption{The scaling dimension $\Delta(k)$ for different $k$. The numerical fitting of $\Delta_N(k)$ follows the form of Eq. \eqref{AdS2_scale}, whereas $\Delta_A(k)$ is evaluated directly using Eq.\eqref{Delta_k}.}
	\begin{tabular}{|c|c|c|c|c|c|c|c|c|c|c|c|c|}
		\hline
		$k$ & 0.920  & 0.930 & 0.940 & 0.960 & 0.970 & 0.980 \\
		\hline
		$\Delta_N(k)$ & 1.15637 & 1.28750 & 1.39300 & 1.56911 & 1.64651 & 1.71899\\
		\hline
		$\Delta_{A}(k)$ & 1.15647 & 1.28764 & 1.39312 & 1.56923 & 1.64662 & 1.71911\\
		\hline
	\end{tabular}
	\label{ta-0}
\end{table}

For the novel metallic phase, characterized by a hyperscaling violation IR geometry, We can easily determine the scaling behaviors of the DC resistivity and thermal conductivity \cite{Donos:2014cya}
\begin{eqnarray}\label{HV_scale}
	\rho \sim T^{- \frac{(1+\gamma)(3-\gamma)}{9+2\gamma+\gamma^2}}\,, \ \ \bar{\kappa}\sim T^{\frac{2(7+4\gamma+\gamma^2)}{9+2\gamma+\gamma^2}}\,.
\end{eqnarray}
As evident, the scaling exponent remains independent of the charge of the axion field $k$. Furthermore, by examining the expression for $\bar{\kappa}$ in Eq.\eqref{HV_scale}, we can see that it tends towards zero as the temperature approaches zero. This behavior is consistent with the findings presented in Fig.\ref{thermal-k_HV}.

\begin{figure}[H]
	\centering
	\includegraphics[width=0.48\textwidth]{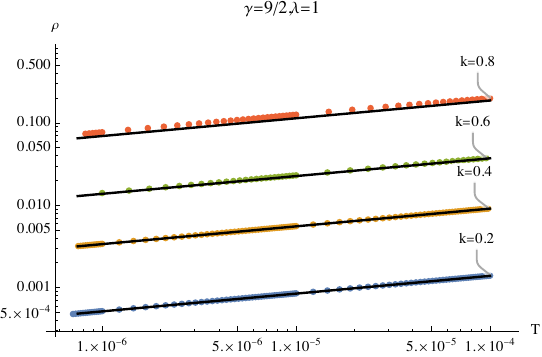}\hspace{0.3mm}
	\includegraphics[width=0.48\textwidth]{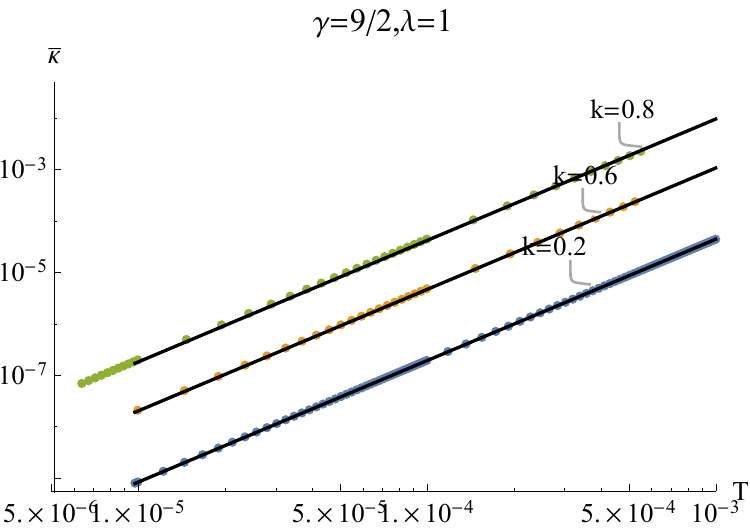}
	\caption{The log-log plot of the DC resistivity $\rho$ and thermal conductivity $\bar{\kappa}$ as a function of temperature $T$ for different $k$ in the novel phase. The numerical results are represented by dots, while the analytical results, evaluated using Eq.\eqref{Delta_k}, are depicted as solid lines.}
	\label{HV_scaling_relationship}
\end{figure}

Fig.\ref{HV_scaling_relationship} shows the log-log plot of the DC resistivity $\rho$ and the DC thermal conductivity $\bar{\kappa}$ as a function of temperature $T$ for different $k$ in the novel phase. It is clearly seen that the numerical results are in good agreement with the analytical ones. 
Quantitatively, we also fit the numerical data with the form of $\rho\sim T^{\alpha}$ and $\bar{\kappa}\sim T^{\beta}$. The fitting results are summarized in Table \ref{ta-1}. On the other hand, since throughout this paper, we fix $\gamma=9/2$, the scaling exponent can be evaluated by Eq.\eqref{HV_scale} as $\alpha\approx 0.21569$ and $\beta\sim 2.36601$, respectively.
Again, the quantitative numerical results are well consistent with the analytical ones determined by Eq.\eqref{HV_scale}.

\begin{table}[H]
	\centering
	\caption{The scaling exponents $\alpha$ and $\beta$ fitted with the form of $\rho\sim T^{\alpha}$ and $\bar{\kappa}\sim T^{\beta}$ in the novel metallic phase.}
	\begin{tabular}{|c|c|c|c|}
		\hline
		$k$ & 0.20  & 0.40 & 0.60   \\
		\hline
		$\alpha$ & 0.21519 & 0.21517 & 0.21394 \\
		\hline
		$\beta$ & 2.36500 & 2.36236 & 2.35810  \\
		\hline
	\end{tabular}
	\label{ta-1}
\end{table}

In this section, we have numerically worked out the scaling behavior of the DC resistivity and the DC thermal conductivity at low temperatures. The numerical results are solidly in agreement with the analytical ones, which are determined by the IR geometry. It also indicated that the numerics we implemented here are robust at extremal low temperatures. The numerical techniques provide us with powerful techniques to deal with the AC conductivities at low temperatures in the next section, which is a hard task to handle especially at extremely low temperatures.

\section{AC conductivity}\label{sec-ac}

In this section, we study the properties of the AC conductivity over the EMDA background along the direction of lattice, i.e., $x$-direction here. To this end, we turn on the following consistent linear perturbations
\begin{eqnarray}\label{linear perturbation}
	g_{tx}=e^{-i \omega t}\delta h_{tx}(z)\,,~~~~~
	A_x=e^{-i \omega t}\delta a_x(z)\,,~~~~~
	\chi=e^{-i \omega t}\delta \chi(z)\,.
\end{eqnarray}
Then we obtain three coupling perturbative equations for $\delta h_{tx}(z)$, $\delta a_x(z)$ and $\delta \chi(z)$.

Without loss of generality, we set $\delta a_x(0)=1$ at the UV boundary ($z=0$), which provides the source of the gauge field.
To guarantee what we extract is the current-current correlator of the dual boundary field theory, we need to impose the boundary condition as $\delta \chi(0)- i k \delta h_{tx}(0)/\omega =0$ at the UV boundary, which comes from the diffeomorphism and gauge transformation.
At the horizon, we shall impose the ingoing boundary conditions. Once the perturbative equations are worked out numerically, we can read off the AC conductivity along $x$-direction by
\begin{eqnarray}\label{optical conductivity}
\sigma(\omega) = \frac{\partial_z \delta a_x (z)}{i \omega \delta a_x (z)} \Big{|}_{z \to 0}\,.
\end{eqnarray}

\begin{figure}[H]
	\centering
	\includegraphics[width=0.32\textwidth]{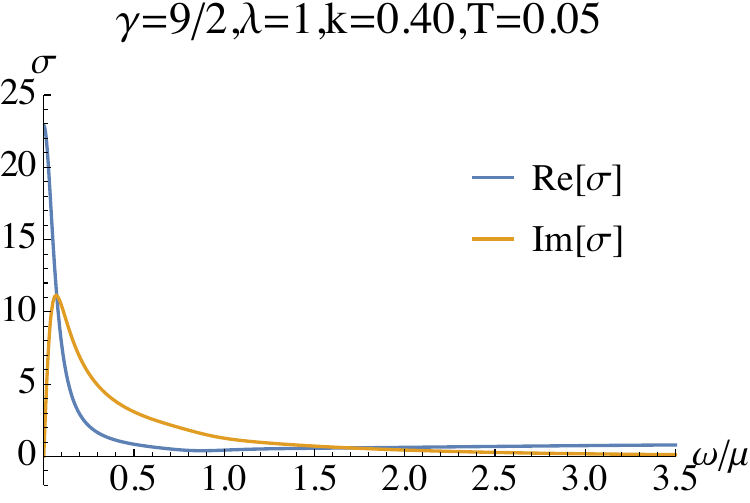}\hspace{0.4mm}	
	\includegraphics[width=0.32\textwidth]{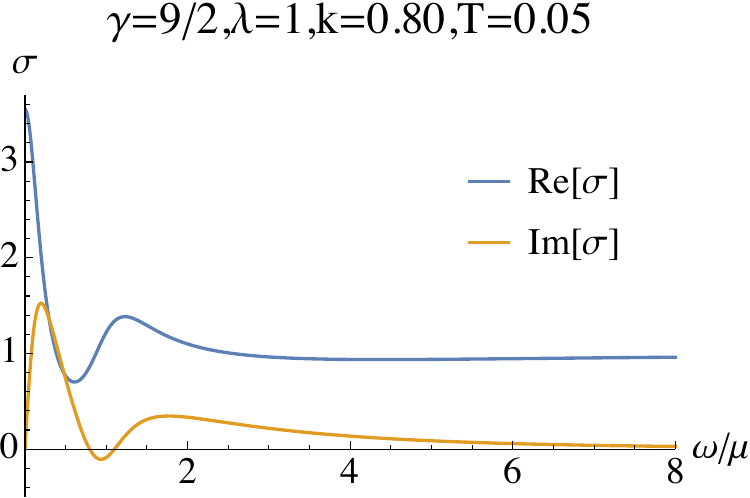}\hspace{0.4mm}
	\includegraphics[width=0.32\textwidth]{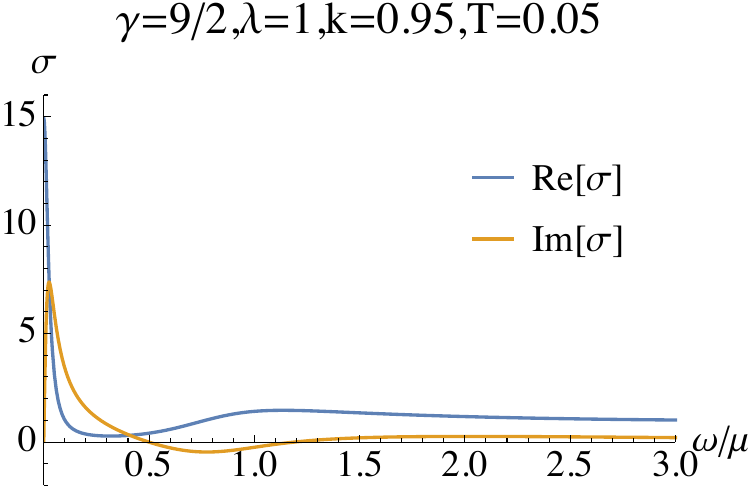}\hspace{0.4mm}	
	\caption{AC conductivity as a function of frequency for different $k = 0.40, 0.80, 0.95$ (from left to right). The temperature is fixed at T=0.05.}
	\label{AC-T0p05}
\end{figure}

Fig.\ref{AC-T0p05} shows the AC conductivity as a function of frequency for different $k$. The conductivity tends to be a constant in the high-frequency limit ($\omega\gg\mu$), which is determined by ultra-violet (UV) AdS$_4$ fixed point. More rich physics lies at the low-frequency region. Moreover, the figure demonstrates Drude-like behavior in the low-frequency range, calling for a deeper dive into that frequency range.

\begin{figure}[H]
	\centering
	\includegraphics[width=0.32\textwidth]{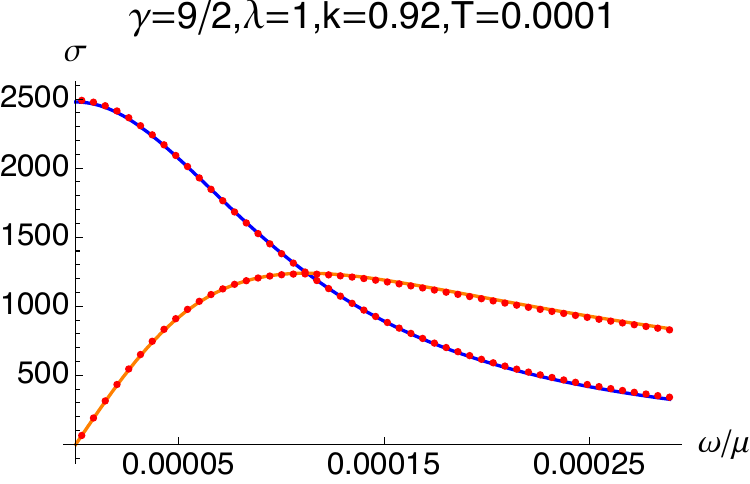}\hspace{0.4mm}	
	\includegraphics[width=0.32\textwidth]{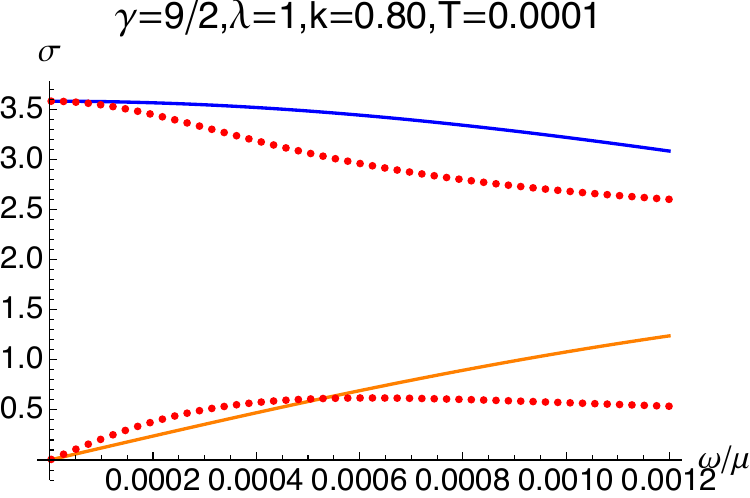}\hspace{0.4mm}	
	\includegraphics[width=0.32\textwidth]{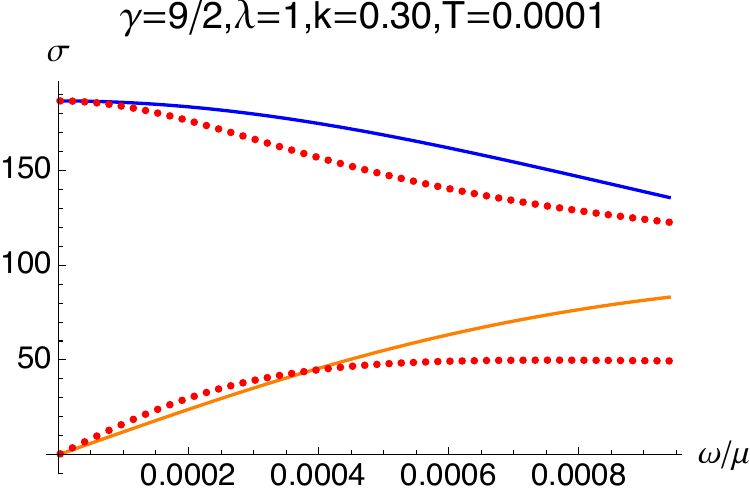}\hspace{0.4mm}	
	\caption{AC conductivity as a function of frequency for different $k = 0.92, 0.80, 0.30$ (from left to right) at $T=0.0001$. The solid lines are the numerical results while the dot lines are fitted by the standard Drude formula \eqref{Drude formula}.}
	\label{fit different k at T 0.0001 Drude}
\end{figure}



To further explore the behavior of the low-frequency AC conductivity, we adopt the numerical fitting to a Drude peak by the Drude formula \eqref{Drude formula} and the modified Drude formula \eqref{Holographic modify formula} in the low-frequency regime. In this process, it is crucial to simultaneously fit the real and imaginary parts of the numerical data. Thus, we perform joint fitting such that we can minimize the residue. 
At the same time, we shall choose a $figure$-$of$-$merit\ function$, also called merit function, to measure the agreement between the numerical data and the model with a particular choice of parameters.
Adjusting the parameters of model to achieve a minimum in the merit function, we can yield the $best$-$fit\ parameters$. Here we consider the chi-square fitting, and the merit function $\bar{\chi}^2$ is given by \cite{article1,article2}
\begin{eqnarray}
	\label{error}
\bar{\chi}^2\equiv \sum_{i=1}^{N} \left(\frac{\sigma_i-\bar{\sigma}_i}{\sigma_s}\right)^2\,,
\end{eqnarray}
{where the $\sigma_s$ is the standard deviation. $\sigma_i$ and $\bar{\sigma}_i$ are the numerical and fitting data, respectively. 

We firstly try to fit the low-frequency AC conductivity at extremal low temperature ($T=0.0001$) by the standard Drude formula \eqref{Drude formula}. As expected, for the normal metallic phase, the AC conductivity at low frequency can be well fitted by the standard Drude formula (the first plot in Fig.\ref{fit different k at T 0.0001 Drude}). However, in the novel metallic phase, the peak is suppressed and becomes flat. It is easy to find that the standard Drude formula is violated (the second and third plots in Fig.\ref{fit different k at T 0.0001 Drude}). This observation prompts us to resort to a modified Drude formula, and we shall again borrow the modified Drude formula \eqref{Holographic modify formula}.
\begin{figure}[H]
	\centering
	\includegraphics[width=0.32\textwidth]{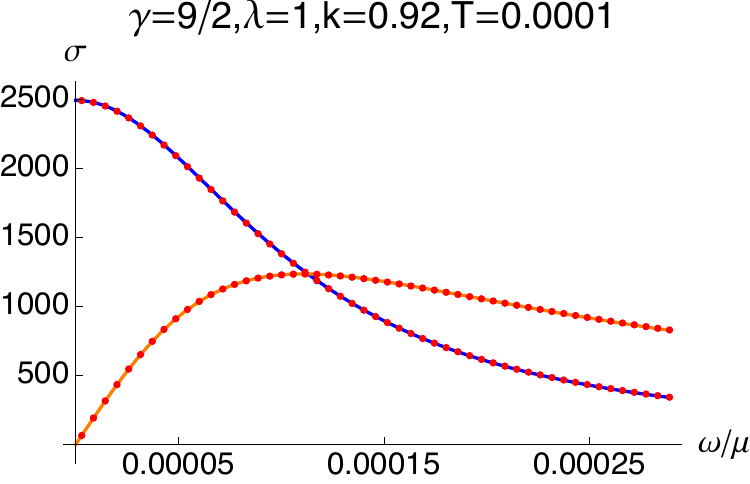}\hspace{0.4mm}
	\includegraphics[width=0.32\textwidth]{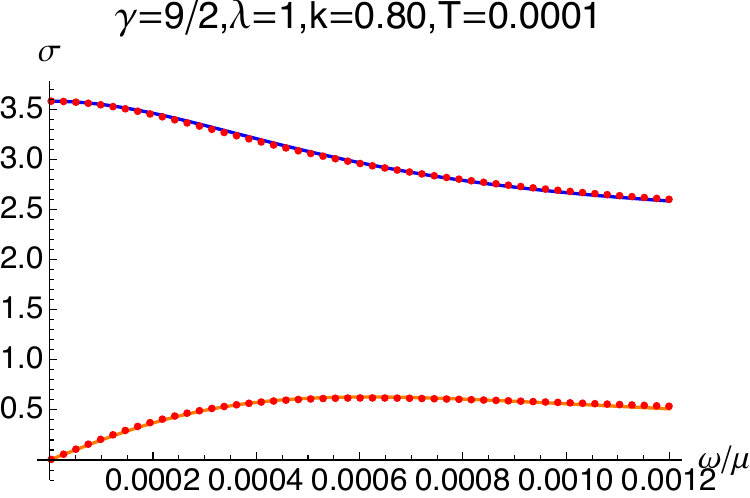}\hspace{0.4mm}	
	\includegraphics[width=0.32\textwidth]{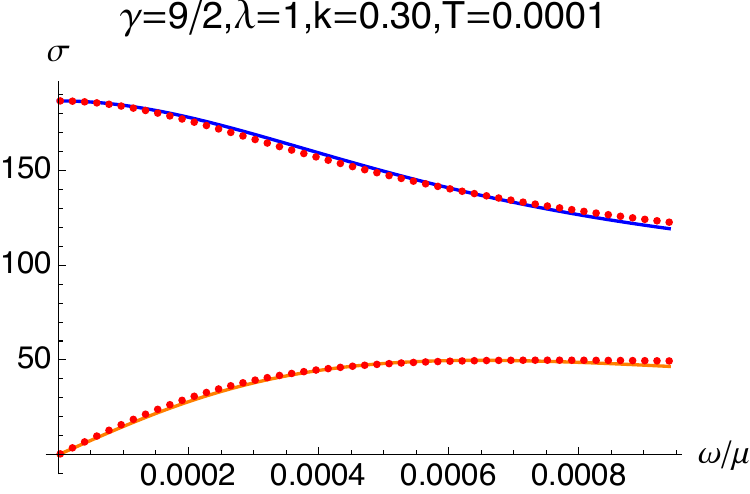}\hspace{0.4mm}	
	\caption{AC conductivity as a function of frequency for different $k = 0.92, 0.80, 0.30$ (from left to right) at $T=0.0001$. The solid lines are the numerical results while the dot lines are fitted by the standard Drude formula \eqref{Holographic modify formula}.}
	\label{fit different k at T 0.0001 Modify}
\end{figure}

\setlength{\tabcolsep}{0.4mm}{
	\begin{table}[H]
		\centering
		\caption{The merit function $\bar{\chi}^2$ as a function of $k$ at $T=0.0001$.}
		\begin{tabular}{|c|c|c|c|c|c|c|c|c|c|c|c|c|}
			\hline
			$k$ & 0.300 & 0.500 & 0.800 & 0.840  & 0.920 & 0.930 & 0.940 \\
			\hline
			$\bar{\chi}^2$ & 3.0334$\times 10^{-6}$ & 1.7835$\times 10^{-6}$ & 1.1565$\times 10^{-6}$ & 1.2791$\times 10^{-7}$ & 8.8007$\times 10^{-10}$ &  2.9743$\times 10^{-14}$ & 7.3914$\times 10^{-17}$\\
			\hline
		\end{tabular}
		\label{ta-4}
	\end{table}
}
\setlength{\tabcolsep}{1 mm}{
\begin{table}[H]
	\centering
	\caption{Fitting parameters $\tau \mu$ and $\sigma_Q/\sigma_{DC}$ for different $k$ at $T=0.0001$.}
	\begin{tabular}{|c|c|c|c|c|c|c|c|c|c|c|c|c|}
		\hline
		$k$ & 0.300 & 0.500 & 0.800  & 0.840  & 0.920  & 0.930 & 0.940 \\
		\hline
		$\tau\mu$ & 1539.21 & 1592.79 & 1626.25 & 1834.11 & 9022.03 & 42929.80 &  175266.43 \\
		\hline
		$\frac{\sigma_Q}{\sigma_{DC}}$ & 0.46622 & 0.52335 & 0.64871 & 0.75673 & 0.00913 & 0.00068 & 0.00005 \\
		\hline
	\end{tabular}
	\label{ta-3}
\end{table}
}

Fig.\ref{fit different k at T 0.0001 Modify} shows the numerical fitting results by the modified Drude formula \eqref{Holographic modify formula}. This fitting works exceptionally well in both metallic phases. Further, we also  quantitatively evaluate this fitting by the merit function \eqref{error}. To this end, we show the merit function $\bar{\chi}^2$ for our fitting in Table \ref{ta-4}. The merit function $\bar{\chi}^2$ is easily seen to be on the order of $10^{-6}$ or even smaller. Therefore, we validate that the modified Drude formula \eqref{Holographic modify formula} does fit well.

Quantitatively, we show the fitting parameters $\tau\mu$ and $\sigma_Q/\sigma_{DC}$ in Table \ref{ta-3}. We would like to emphasize that for a fixed $ \sigma_{DC} $, a larger $ \sigma_Q $ indicates a more incoherent system. We measure this coherence with $\sigma_Q/\sigma_{DC}$, and the more this value goes up, the less coherent the system is.
Then, we summarize the main characteristics of the low-frequency AC conductivity at extremal low temperatures as follows.
\begin{itemize}
	\item The normal metallic phase is the coherent phase, which is validated by $\sigma_Q/\sigma_{DC}\ll 1$ (the last three ones in Table \ref{ta-3}). What is particularly interesting is that $\sigma_Q/\sigma_{DC}$ rapidly approaches to be vanishing as the strength of the momentum dissipation enhances. It suggests that the coherent behavior is independent of the momentum dissipation. This phenomenon is completely different from the usual EMA model or Gurbser-Rocha-axions model, where the momentum dissipation induces a transition from coherent phase to incoherent one \cite{Davison:2015bea,Zhou:2015qui,Andrade:2013gsa,Davison:2013jba,Jeong:2018tua,Jeong:2021wiu}.
	\item The novel metallic phase has a non-vanishing $\sigma_Q/\sigma_{DC}$ , which suggests an incoherent behavior. In this phase, the incoherent behavior becomes stronger when the strength of the momentum dissipation grows. This observation is consistent with that in most of the holographic models \cite{Davison:2015bea,Zhou:2015qui,Andrade:2013gsa,Davison:2013jba,Jeong:2018tua,Jeong:2021wiu}.
    \item Whether in the normal metallic phase or in the novel metallic one, the relaxation time increases as the momentum dissipation enhances, which is consistent with that of the usual holographic axions model \cite{Andrade:2013gsa,Wu:2018zdc}.
\end{itemize}

In conclusion, the low-frequency AC conductivities of the normal metallic phase and the novel metallic phase differ noticeably. These differences are due to the two phases' differing IR geometries. 
We expect to reveal the mechanisms underlying these characteristics of AC conductivity in the near future. 
There are at least two ways to accomplish this. Following the method in \cite{Davison:2015bea,Zhou:2015qui}, we can test the robustness of the modified Drude formula \eqref{Holographic modify formula}. Especially, it may address what determines the relaxation time and the intrinsic conductivity in this model. However, in order to deduce the low-frequency AC conductivity behavior, we must first decouple the linearized perturbative equations, which is a difficult task in our model.
We can refer to the work \cite{Zhou:2015qui} for detailed discussions. Another approach for caculating low-frequency AC conductivity analytically is the matching method, which has been widely used in holography (see for example, \cite{Faulkner:2009wj,Iqbal:2011ae,Wu:2013xta,Wu:2013vma,Kuang:2015mlf,Davison:2013jba,Liu:2012tr,Edalati:2009bi,Cai:2009zn}).

Furthermore, the momentum relaxation results in not only finite DC electric conductivity but also finite DC thermal conductivity. However, it is challenging to establish a correlation between the temperature-dependent behavior of DC electric and thermal conductivity and the coherent or incoherent behavior of AC conductivity. As a result, investigating the AC thermal conductivity in our current model would be fascinating since it would help us elucidate the relationship between DC and AC transports, and explore the coherent or incoherent nature of AC transport.

\section{Conclusion and discussion}\label{sec-con}

In this paper, we investigate the properties of charge transport on a special EMDA theory with novel QPTs \cite{Donos:2014uba,Fu:2022qtz}. We specially focus on the novel metallic state, which has two distinct metallic phases: normal metallic phase and  novel metallic phase \cite{Donos:2014uba,Fu:2022qtz}. We investigate the scaling behaviour of two different metallic phases using numerical approaches. The analytical results determined by the near horizon geometry are in great agreement with the numerical ones \cite{Donos:2014uba,Donos:2014cya}.

We systematically investigate the low-frequency AC conductivity characteristics in the two distinct metallic phases since they better depict the exotic metallic behavior. We postulate that the modified Drude formula \eqref{Holographic modify formula}, which was obtained in the holographic dual system of EMA theory and Gubser-Rocha-axions model \cite{Davison:2015bea,Zhou:2015qui}, is universal and also applicable to our current novel metallic states. The study discovers that the normal metallic phase is a coherent system, which is a notable property and distinct from the dual system of typical axions model which transits from coherent phase to incoherent phase as the momentum dissipation increases, whereas the novel metallic phase is an incoherent system. When the strength of the momentum dissipation increases, the incoherent behavior in the novel metallic phase becomes stronger. The disparities between the two phases can be attributed to the differing IR geometries. We expect to reveal the mechanisms underlying these characteristics of AC conductivity in the near future.

The special EMDA model proposed in \cite{Donos:2014uba,Fu:2022qtz} provided a platform for studying the QPT. There are several avenues that should be pursued further. To begin, we may look at quantum information measurement in this EMDA model, such as the holographic entanglement entropy and the entanglement wedge minimum cross-section, as proposed in \cite{Ling:2015dma,Ling:2016wyr,Ling:2016dck,Jeong:2022zea}. It is also important to investigate the properties of AC conductivity in other states, particularly the insulating phases reported in \cite{Fu:2022qtz}. Furthermore, we would want to build an anisotropic 
background based on this EMDA model to investigate its dynamical features. Such dynamical features have been investigated in the anisotropic background based on the Q-lattice model \cite{Liu:2021stu}. It is reasonable to expect that introducing anisotropy will result in more strange events in the existing EMDA model. It will also be interesting to investigate the holographic fermionic spectral functions over this background, which may be used to identify distinct holographic phases, such as \cite{Fang:2015dia,Ling:2014bda,Alsup:2014uca,Ling:2016ewj,Jeong:2019zab}.

\acknowledgments

This work is supported by the Natural Science Foundation of China under Grant Nos. 11905083, 11775036, 12147209, the Postgraduate Research \& Practice Innovation Program of Jiangsu Province under Grant Nos. KYCX20\_2973 and KYCX22\_3451, Fok Ying Tung Education Foundation under Grant No. 171006, Natural Science Foundation of Jiangsu Province under Grant No.BK20211601, Top Talent Support Program from Yangzhou University and the Science and Technology Planning Project of Guangzhou (202201010655).

\bibliographystyle{style1}
\bibliography{Ref}
\end{document}